\documentclass[aps,prl,twocolumn,showpacs,preprintnumbers,amsmath,amssymb]{revtex4-1}
\usepackage{latexsym}
\usepackage{graphicx}
\usepackage{graphics}
\usepackage[T1]{fontenc}
\usepackage{multirow}

\begin{document}

\author{Panagiotis Vergyris$^{1}$, Charles Babin$^2$, Raphael Nold$^2$, Elie Gouzien$^1$, Harald Herrmann$^{3}$, Christine Silberhorn$^3$, Olivier Alibart$^1$, S\'ebastien Tanzilli$^1$}
\author{Florian Kaiser$^{1,2}$}
\email{f.kaiser@pi3.uni-stuttgart.de}

\affiliation{$^1$Universit\'e C\^ote d'Azur, Institut de Physique de Nice (INPHYNI), CNRS, Parc Valrose, 06108 Nice Cedex 2, France\\
$^2$3rd Institute of Physics, IQST, and Research Center SCoPE, University of Stuttgart, 70569 Stuttgart, Germany\\
$^3$Integrated Quantum Optics, Universit\"at Paderborn, Warburger Strasse 100, 33098 Paderborn, Germany}

\title{Two-photon phase-sensing with single-photon detection}

\begin{abstract}
Path-entangled multi-photon states allow optical phase-sensing beyond the shot-noise limit, provided that an efficient parity measurement can be implemented. Realising this experimentally is technologically demanding, as it requires coincident single-photon detection proportional to the number of photons involved, which represents a severe challenge for achieving a practical quantum advantage over classical methods.
Here, we exploit advanced quantum state engineering based on superposing two photon-pair creation events to realise a new approach that bypasses this issue. In particular, optical phase shifts are probed with a two-photon quantum state whose information is subsequently effectively transferred to a single-photon state. Notably, without any multiphoton detection, we infer phase shifts by measuring the average intensity of the single-photon beam on a photodiode, in analogy to standard classical measurements. Importantly, our approach maintains the quantum advantage: twice as many interference fringes are observed for the same phase shift, corresponding to $N=2$ path-entangled photons.
Our results demonstrate that the advantages of quantum-enhanced phase-sensing can be fully exploited in standard intensity measurements, paving the way towards resource-efficient and practical quantum optical metrology.
\end{abstract}

\keywords{Quantum Metrology, Entanglement, Guided-Wave Optics, Photonics, phase-sensing, Quantum physics}
\maketitle

Exploiting quantum coherence and correlations allows performing phase-sensitive measurements with a precision surpassing classical approaches which are ultimately shot-noise limited.
For optical phase-sensing, the textbook example considers a Mach-Zehnder interferometer in which a phase shift is to be detected using $N$ probe photons.
The optimal strategy employs a path-entangled $N00N$-state, representing a coherent superposition of having $N$ photons in one interferometer arm with zero in the other, and vice versa.
In this case, considering even and odd parity photonic states at the interferometer output, leads to interference fringes that run $N$ times faster compared to classical approaches based on non-entangled photons.
As a consequence, the phase-sensing precision $\Delta \phi$ is improved from $\Delta \phi \sim \tfrac{1}{\sqrt{N}}$ (classical shot-noise limit) to $\Delta \phi \sim \tfrac{1}{N}$ (quantum Heisenberg limit)~\cite{Giovannetti_QuMetReview_2004,Dowling_N00N_2008}.

However, the price to pay for this quantum feature is significant. In previous approaches, photonic parity detection (even or odd number of photons at each interferometer output) required simultaneous detection of multiple photons of the order of $N$~\cite{Walther_4photon_2004,Mitchell_superresolution_2004,Nagata_superresolution_2007,Higgins_entanglement-free_2007,Matthews_multiphoton-wg_2009,Afek_high-noon_2010,Crepsi_protein-concentration_2012,Ono_microscopy_2013,Israel_microscopy_2014,Wolfgramm_Delicate_2013, Slussarenko_2017,Cimini_PRA_2019}. This circumstance results in excessive measurement times compared to classical approaches.

In this article, we introduce and experimentally demonstrate a new scheme for parity detection of an arbitrary $N00N$-state based on simple (and fast) single-photon detection.
The detection scheme makes our approach particularly interesting for quantum optical metrology applications in which a measurement signal is usually averaged over a given integration time to improve signal-to-noise. As our detection scheme resembles closely the classical one, this averaging can be conveniently done by low-pass filtering the output signal of a standard photodiode (without single-photon sensitivity), thus resolving the long-standing parity detection issue in quantum optical metrology.
In addition, by eliminating the necessity for single- and multi-photon coincidence detection, our approach can take advantage of high-power photon pair sources with fluxes on the order of microwatts~\cite{Dayan_bright_2004,Avi_bright_2005,Dayan_bright_2005}. This promises significant reduction of measurement times, opening new avenues in addressing quantum-enhanced optical sensing applications.

\begin{figure}
\includegraphics[width=1\columnwidth]{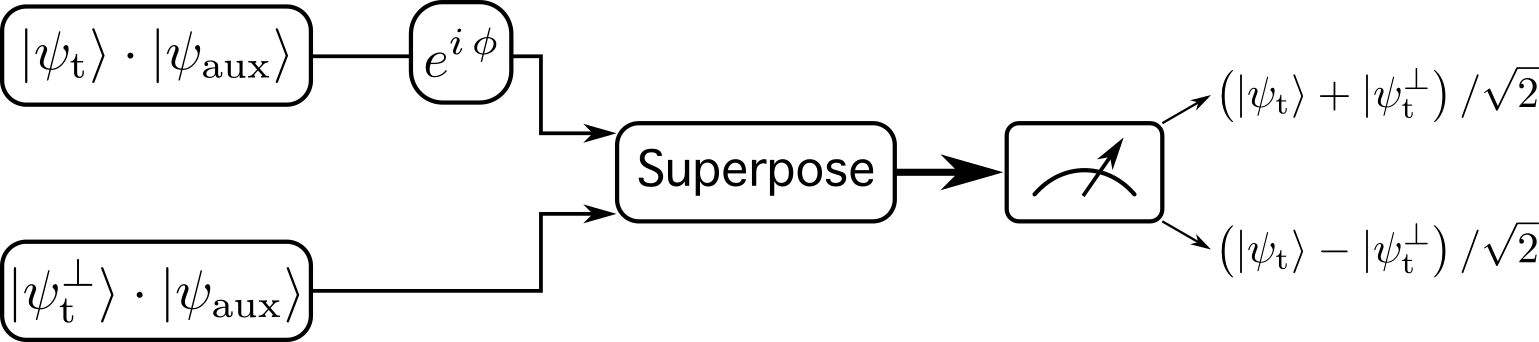}
\caption{\textbf{Schematic description of the sensing scheme.} Two multi-photon quantum state contributions are superposed. Each state comprises of $N-1$ auxiliary single-photon states summarised within $|\psi_{\rm aux} \rangle = \sqrt{p} \Pi_{k=1}^{N-1} |\psi_k \rangle$. Each contribution is associated with a target mode (subscript t). The target modes ($|\psi_{\rm t} \rangle$ and $|\psi^{\dagger}_{\rm t} \rangle$) of both contributions are orthogonal. Only one of the two contributions is subjected to the phase shift. After superposition, the entire $N$-photon phase shift information can be retrieved by a measurement on the target mode photon. All auxiliary photons are irrelevant for retrieving phase information.\label{Sketch}}
\end{figure}

In the following, we first give a brief theoretical description of our approach, followed by detailing the experimental realisation.
As shown in the schematic diagram in \figurename~\ref{Sketch}, the underlying principle of our strategy is based on quantum state engineering using induced coherence without induced emission~\cite{Mandel_induced1_1991,Mandel_induced2_1991,Lemos_Q-imaging_2014,Kalashnikov_IR_2016}, i.e. we generate a photon number state in a superposition of being generated at two different locations.
Assume that the photon state contribution from the first location is:
\begin{equation}
|\Psi_1 \rangle = \sqrt{p} |\psi_{\rm t} \rangle \Pi_{k=1}^{N-1} |\psi_k \rangle.
\end{equation}
Here, $p < 1$ stands as the probability to generate the contribution, in which each $|\psi \rangle$ denotes a single photon state, $\rm t$ stands for the target photon mode (from which we want to extract the multi-photon phase information in the end), and all $k$ are axillary photon modes. Note that we require that all modes $k$ are distinguishable from the target mode $\rm t$, e.g. by their wavelengths, however the modes $k$ do not have to be distinguishable in-between themselves.\\
In the second step, $|\Psi_1 \rangle$ is made to interact with a phase object, that induces a phase shift on each individual photon. All phase shifts together sum up to a total $N$-photon phase shift:
 \begin{equation}
\phi = \phi_{\rm t} + \sum_{k=1}^{N-1} \phi_k.
\end{equation}
The phase object transforms $|\Psi_1 \rangle$ to $e^{i \, \phi} |\Psi_1 \rangle$.\\
The third step is to coherently superpose $e^{i \, \phi} |\Psi_1 \rangle$ with the photonic contribution from the second location:
\begin{equation}
|\Psi_2 \rangle = \sqrt{p} |\psi^{\perp}_{\rm t} \rangle \Pi_{k=1}^{N-1} |\psi_k \rangle.
\end{equation}
Here, $|\psi^{\perp}_{\rm t} \rangle$ stands for a photonic mode that is orthogonal to $|\psi_{\rm t} \rangle$, i.e. $\langle \psi^{\perp}_{\rm t} | \psi_{\rm t} \rangle = 0$, e.g. $| \psi_{\rm t} \rangle$ and $| \psi^{\perp}_{\rm t} \rangle$ can represent orthogonal polarisation modes. The overall superposition state reads now $e^{i\,\phi}\sqrt{p} |\psi_{\rm t} \rangle \Pi_{k=1}^{N-1} |\psi_k \rangle + \sqrt{p} |\psi^{\perp}_{\rm t} \rangle \Pi_{k=1}^{N-1} |\psi_k \rangle$, which can be further simplified to:
\begin{equation}
\left( e^{i\,\phi} |\psi_{\rm t} \rangle + |\psi^{\perp}_{\rm t} \rangle\right)  |\Psi_{\rm aux} \rangle. \label{Important_Equation}
\end{equation}
Here, $|\Psi_{\rm aux} \rangle = \sqrt{p}  \Pi_{k=1}^{N-1} |\psi_k \rangle$ comprises all auxiliary modes $k$. Importantly, equation~\ref{Important_Equation} shows that the $N$-photon phase shift is now observed on the target single-photon mode. To extract the full phase shift information, we can now ignore the auxiliary photon modes and solely measure the target mode in the phase-sensitive basis of the vectors $\left(|\psi_{\rm t} \rangle \pm |\psi^{\perp}_{\rm t}\right)/\sqrt{2}$.

\begin{figure}
\includegraphics[width=1\columnwidth]{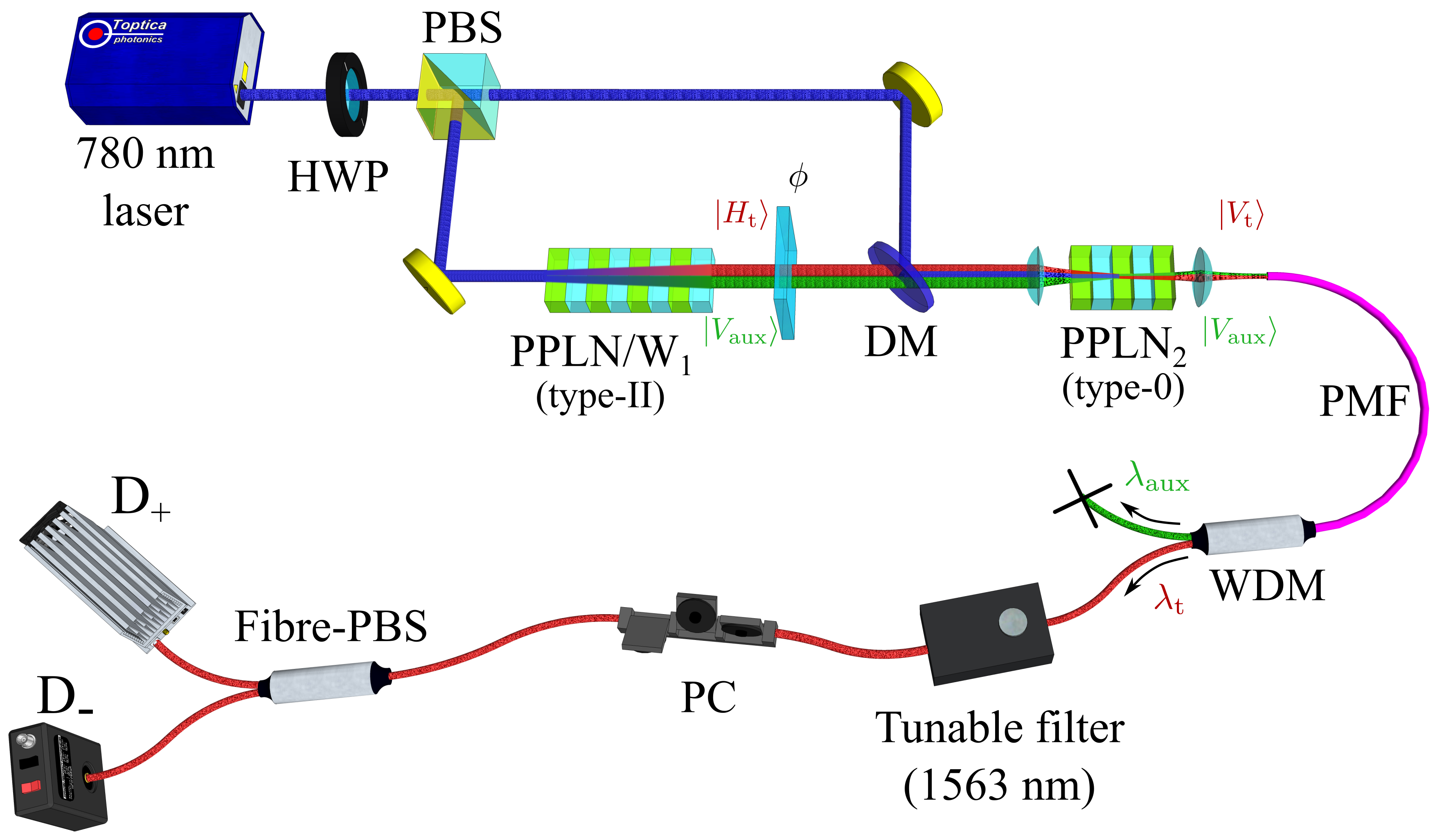}
\caption{\textbf{Detailed experimental setup.} 780\,nm pump laser light (blue lines) is split at a PBS and pumps one type-II PPLN/W$_1$ and one type-0 PPLN$_2$ towards the generation of a delocalized photon pair (red and green lines). A glass plate in the lower arm of the interferometer induces a two-photon phase shift to the pair contribution coming from PPLN/W$_1$. The contribution is coherently overlaped with the contribution of PPLN$_2$ and further coupled into a single-mode PMF (purple line) whose birefringence compensates the birefringence of PPLN/W$_1$. The auxiliary photon at $\lambda_{\rm aux}$ is discarded by a WDM, and the target photon at $\lambda_{\rm t} = 1563\rm\,nm$ is further spectrally filtered. Its polarisation state is then projected into the diagonal basis using a PC, a fibre-PBS and an intensity detector (D$_{\pm}$).\label{Setup2bis}}
\end{figure}

In the following, we detail the experimental realisation of the protocol. For our proof-of-concept demonstration, we use two-photon states ($N=2$), created in via spontaneous parametric downconversion (SPDC) in nonlinear crystals. Compared to previous approaches that exploited superposition of multiple two-photon states~\cite{Lemos_Q-imaging_2014}, and SU(1,1) interferometers~\cite{Chekhova_Nonlinear_2016}, our approach is based on two different nonlinear crystals. Crucially, this enables the demonstration of the aforementioned phase sensing protocol. 
The experimental setup is shown in \figurename~\ref{Setup2bis}.
A continuous-wave 780\,nm pump laser is sent to a polarising beam-splitter (PBS), defining the input of the Mach-Zehnder type interferometer.  A half-wave plate (HWP) in front of the PBS allows adjusting the optical power in both interferometer paths. 
In the lower path, a 3.2\,cm long periodically poled lithium niobate waveguide (PPLN/W$_1$) is employed that is optimised for type-II spontaneous parametric down-conversion (SPDC).
Via this process, we generate the first multi-photon state contribution:
\begin{equation}
|\Psi_{\rm type-II} \rangle = \sqrt{1-p} \, |{\rm vac} \rangle + \sqrt{p} \, |H_{\rm t} \rangle |V_{\rm aux} \rangle.
\end{equation}
Here, $|\rm vac \rangle$ denotes the vacuum state, and $p < 1$ is the probability of generating a photon pair per single-photon coherence time (5\,ps or 150\,ps, depending on the measurement).
Here, $H$ and $V$ denote horizontal and vertical polarisation modes, and the subscripts t and aux denote the target and auxiliary modes. To ensure that those modes are distinguishable, we choose a non-degenerate phase matching, such that the target mode is created at a wavelength $\lambda_{\rm t} = 1563\rm\,nm$ and the auxiliary mode at $\lambda_{\rm aux} = 1557\rm\,nm$, respectively.
The cross-polarised pair contribution is subsequently sent through a phase object, here a glass plate of thickness $d=1\rm\,mm$ and refractive index $n \approx 1.444$. In analogy to standard quantum metrology based on $N00N$-states, this leads to a two-photon phase shift of $\phi = 2\,\pi\,n\,d \cdot\left(\lambda_{\rm t}^{-1} + \lambda_{\rm aux}^{-1} \right) \approx 4\,\pi\,n\,d \cdot\lambda_{\rm t}^{-1}$. The phase shift is to be compared to the light propagating in the upper interferometer arm.
In order to superpose the cross-polarised two-photon state with another one, we recombine it with the pump laser into the same spatial mode at the output of the interferometer using a dichroic mirror (DM). Then, all light fields are focussed in a 5\,mm long periodically poled lithium niobate bulk crystal (PPLN$_2$), where pump photons can be converted to a co-polarised photon pair contribution through type-0 SPDC:
\begin{equation}
|\Psi_{\rm type-0} \rangle = \sqrt{1-p} \, |{\rm vac} \rangle + \sqrt{p} \, |V_{\rm t} \rangle |V_{\rm aux} \rangle.
\end{equation}
Light is collected into a 3.2\,m long polarisation maintaining (PM) fibre whose birefringence compensates the birefringence of both nonlinear crystals~\cite{Kaiser_typeII_2012}. Light detection effectively rejects the vacuum contribution, such that we obtain the desired quantum state:
\begin{equation}
\left( {\rm e}^{{i}\,\phi} |H_{\rm t} \rangle + |V_{\rm t} \rangle \right)  |V_{\rm aux} \rangle. \label{PhaseTransfer}
\end{equation}
We mention again that equation~\ref{PhaseTransfer} shows that the superposition of the emissions of PPLN/W$_1$ and PPLN$_2$ effectively transfers the entire two-photon phase shift onto the photon at $\lambda_{\rm t}$. $|V_{\rm aux} \rangle$ is factored out and does neither contain information about the phase shift nor about the origin of the photon pair. Therefore, it does not need to be detected, and we take it out of its partner photon using a fibre wavelength division multiplexer (WDM). For signal photons at $\lambda_{\rm t}$ we employ an additional bandpass filter (55\,pm or 1000\,pm bandwidth) in order to ensure good spectral mode overlap from both downconversion sources.
Then, a fibre polarisation controller (PC) is used to rotate the polarisation state of the photons at $\lambda_{\rm t}$ to the phase-sensitive diagonal basis (\textit{i.e.} by $45^{\circ}$).
Quantum state projection is performed using a fibre-PBS after which light intensities $I_{\pm}$ in the upper/lower detectors D$_{\pm}$ follow:
\begin{equation}
I_{\pm} \propto 1 \pm \mathcal{V}\cdot \cos \phi. \label{IntensityEquation}
\end{equation}
Here, $\mathcal{V}$ stands as the interference pattern visibility which takes into account experimental imperfections such as optical losses and non-ideal mode matching (for more details, see Supplementary Note 3).
Depending on the measurement, we use either an InGaAs SPD, featuring 15\% detection efficiency and 360 dark counts per second (D$_+$; IDQuantique id230), or a standard photodiode (D$_-$; Newport model 2153).

\begin{figure}
\includegraphics[width=1\columnwidth]{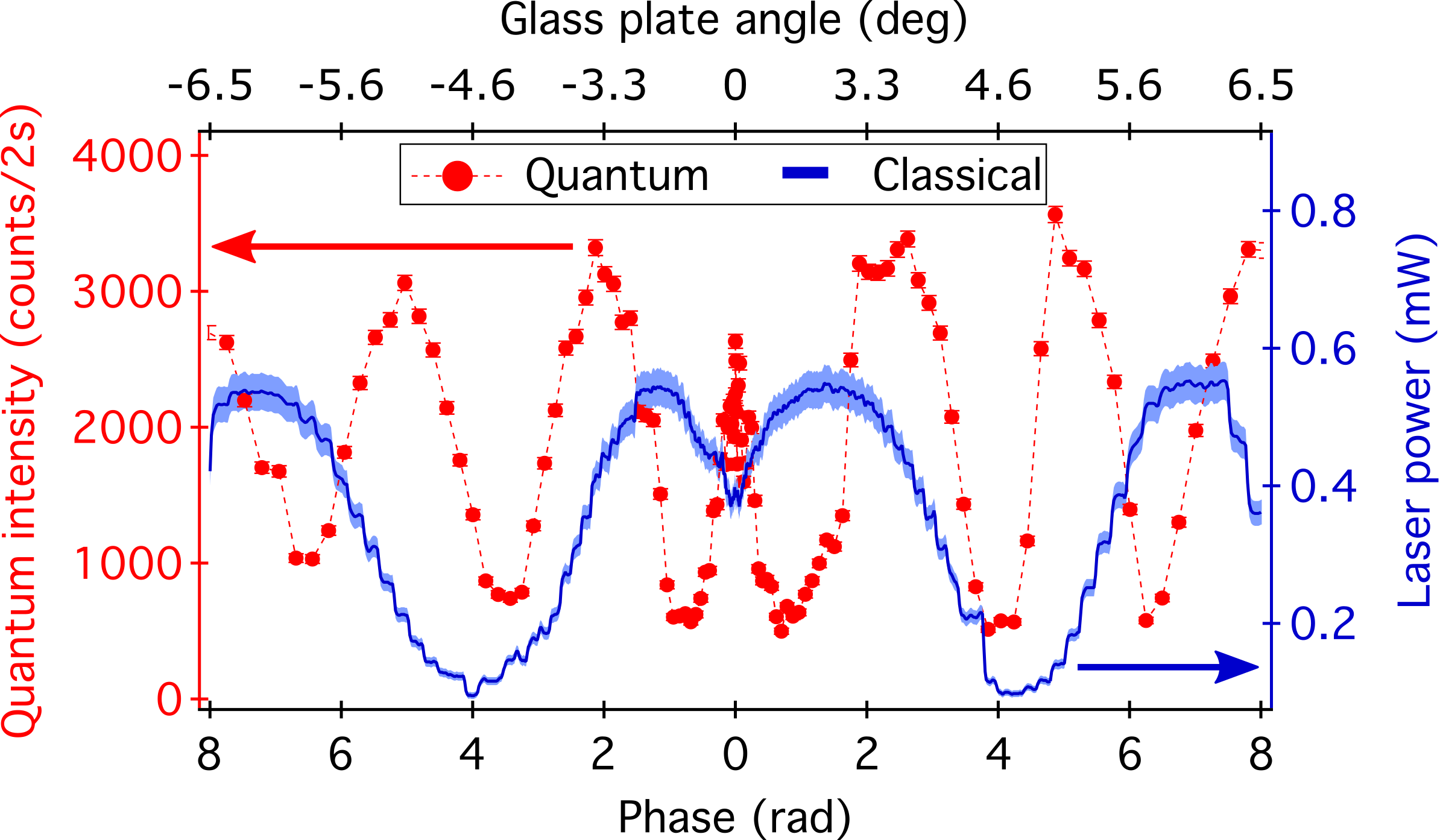}
\caption{\textbf{Optical phase shift measurements.} Blue lines and red dots represent classical and quantum data, respectively. Red dashed lines are a guide to the eye. The top $x$-axis represents the glass plate angle in degrees, while the bottom $x$-axis shows the corresponding phase shift. The quantum-enhanced strategy leads to a two-fold increased phase sensitivity. Notably, in both cases classical light intensities are measured. For the quantum-enhanced data, the detectors noise contribution of 360 counts per second has been subtracted. The shaded blue area represents the error bar of the classical measurement. For the quantum-enhanced data, standard error bars are show. Error bars correspond to one standard error. Throughout the measurement, the interferometer was not actively stabilised, such that occasional phase kicks and small drifts are ovserved.\label{Results1bis}}
\end{figure}

Our first experimental results intend to demonstrate that the detected single photons at 1563\,nm actually carry the phase shift information of both photons in each pair. To this end, we benchmark our approach against a standard Mach-Zehnder interferometer, illuminated by a 1560\,nm continuous-wave laser.
In both cases, we induce phase shifts by tilting the glass-plate and record intensity interference fringes.
The blue curve in \figurename~\ref{Results1bis} shows the fringes obtained with the 1560\,nm laser. The red dots show the results obtained with the quantum-enhanced protocol. For the latter, we use detector D$_+$ and the 55\,pm bandpass filter.
The symmetry point of the interferograms allows to define the angle of normal light incidence on the glass plate.
The results show that the quantum-enhanced method leads to the observation of twice as many fringes per phase interval, proving that the target single-photon mode at $\lambda_{\rm t}=1563\rm\,nm$ actually carries the phase shift accumulated by the pair contribution.

Reduced fringe visibilities at negative tilting angles are explained by an angle-dependent beam-displacement, leading to reduced mode overlap in PPLN$_2$. In fact, the experimental setup has been optimised for angles around $5^{\circ}$ where we obtain $\mathcal{V} = 77 \pm 3\%$, in good agreement with the theoretically expected value of $\mathcal{V}_{\rm theo} = 79\%$ (for more details see Supplementary Note 2). We mention also that the interferometer was not actively stabilised, such that occasional phase kicks and slow drifts are present.

At this point, we stress two experimental key assets of our approach.\\
First, our approach is different from previous quantum sensing schemes in which intensity detection was used, however only classical-like single-photon phase sensitivity was demonstrated~\cite{Lemos_Q-imaging_2014}. Other schemes in which two-photon sensitivity was observed relied always on multi-photon and/or coincidence detection~\cite{Walther_4photon_2004,Mitchell_superresolution_2004,Nagata_superresolution_2007,Higgins_entanglement-free_2007,Matthews_multiphoton-wg_2009,Afek_high-noon_2010,Crepsi_protein-concentration_2012,Ono_microscopy_2013,Israel_microscopy_2014,Wolfgramm_Delicate_2013, Slussarenko_2017,Cimini_PRA_2019}.\\
And second, the same phase-shift could be obtained by transmitting the short-wavelength pump laser through the phase object. However, it was shown that chemical reactions, biological samples, and atomic spin ensembles do get perturbed when exposed to short-wavelength light, which is why quantum sensing at longer wavelengths is to be preferred~\cite{Wolfgramm_Delicate_2013,Cimini_PRA_2019}.

\begin{figure}
\includegraphics[width=0.8\columnwidth]{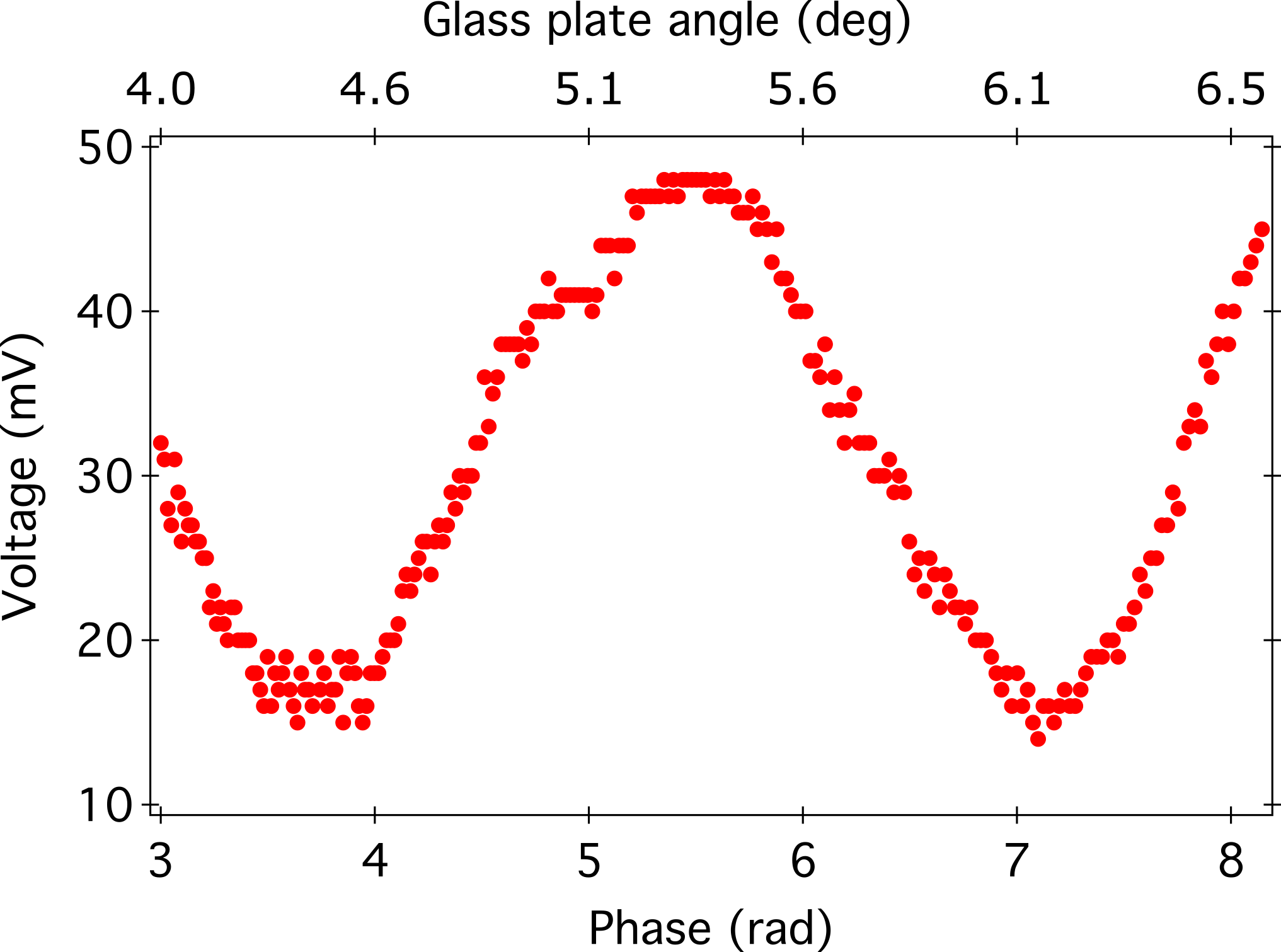}
\caption{\textbf{Quantum results with a classical photodiode.} Interference fringes with half the periodicity of the classical approach are maintained in standard intensity measurements, proving that the quantum advantages of $N00N$-states can be fully exploited without the need for PNRDs and coincidence detection. Throughout the measurement, the interferometer was not actively stabilised, such that occasional phase kicks and small drifts are ovserved.\label{Results2bis}}
\end{figure}

As our second experiment results, we demonstrate quantum-enhanced phase-sensing with light detection using a standard photo-diode (D$_-$), i.e. without single-photon sensitivity. Due to the rather low SPDC efficiency of PPLN$_2$, we leverage the signal by increasing the spectral bandwidth of the filtered photons to 1\,nm. Additionally, the photodiode amplifier's noise is suppressed by employing a lock-in detection scheme. For this, the pump laser was chopped at 600 Hz with a duty cycle of 50\%. The photodiode signal was mixed with the 600 Hz modulation signal and low-pass filtered with a time constant of 3\,s.
The related experimental results are shown in \figurename~\ref{Results2bis}.
As for the previous measurements, intensity interference fringes are obtained as a function of the glass plate angle. The fringe spacing is still half as short as for the standard method using a laser beam, therefore proving that the advantages of $N00N$-states for phase-sensing can be fully exploited using standard intensity measurements.

We note that the fringe visibility in the photodiode measurements is reduced to $\sim$52\%. However, this is not due to some fundamental limitation. Here, visibilities are almost exclusively limited by the rather low photon-pair generation rate of PPLN$_2$, delivering only a few 100\,fW of photon pair beam intensity within 1\,nm spectral bandwidth. This is only slightly above the noise level of the photodiode amplifier and therefore explains the observation of reduced visibilities (detector noise equivalent power $\approx \rm\,23 fW/\sqrt{Hz}$).

This limitation can be overcome with higher photon pair beam intensities in the microwatt regime~\cite{Dayan_bright_2004,Avi_bright_2005,Dayan_bright_2005,Boitier_bright_2011}, for which we expect near optimal fringe visibilities, even without the use of a lock-in detection scheme.
Thus, by employing two photodiodes as detectors D$_{\pm}$, and subtracting both photo currents, we expect to perform quantum-enhanced phase sensing below the shot-noise limit.
Interestingly, our theoretical modelling shows that the two-fold increased number of interference fringes persists even when generating squeezed states of light at much higher pump powers (for more details, see Supplementary Note 1).

In summary, we have introduced and demonstrated a novel strategy to exploit the phase-sensitivity advantage of two-photon $N00N$-states in standard intensity measurements.

Importantly, the visibility of the interference fringes is only limited by the optical loss between the two SPDC sources (see Supplementary Note 1). In order to obtain an exploitable quantum advantage, the following relation must hold~\cite{Slussarenko_2017}: $\mathcal{V}^2 \,N > 1$.
Note that no requirements are set on the detector's quantum efficiency, as our approach permits using the same light intensity detectors as in classical phase-sensing.

Due to the high quantum efficiencies and saturation levels of standard photodiodes, future quantum optical sensing applications should benefit from a tremendous measurement speed-up compared to state-of-the-art schemes. Possible applications may cover quantum-enhanced phase-contrast microscopy~\cite{Ono_microscopy_2013}, and measurement of material parameters, such as chromatic dispersion~\cite{Kaiser_CD_2017}.

From a fundamental point of view, it will be interesting to study whether additional phase information can be extracted of the auxiliary photon mode, similarly to previous work~\cite{Jin_2013}.
Ideally, this should be implemented while still maintaining coincidence-free detection.

In order to further improve the phase sensitivity, our scheme could be extended to higher photon numbers, for example by coherently superposing the emissions of two photonic triplet sources~\cite{Huebel_triplets_2010,Krapick_triplets_2016,Moebius_triplets_2016}. Another approach could be based on generating high photon number states using several single-photon emitters. The challenge with the latter approach is that near-unity light collection efficiency is required to guarantee the exact photon number. Any deviation from the desired photon number will result in reduced fringe visibilities.
Alternatively, it would be interesting to study whether \textit{approximate} photon number states, generated by superposing coherent and quantum light sources, could benefit from our scheme~\cite{Afek_high-noon_2010}.

We therefore believe that our scheme will have a significant impact in the field of quantum information science, covering fundamental and applied aspects of quantum control and metrology.

\section{Supplementary Material}
Two supplementary notes are provided, giving additional details on the transition from our low photon number experimental demonstration to squeezed states of light. Additionally, we provide the theoretical framework to compute interference fringe visibility loss as a function of the power transmission coefficient between the two nonlinear crystals.

\section{Acknowledgements}

P.V., C.B., E.G., O.A., S.T., and F.K. acknowledge financial support from the Foundation Simone \& Cino Del Duca, l'Agence Nationale de la Recherche (ANR) for the e-QUANET, CONNEQT, INQCA, and SPOCQ projects (grants ANR-09-BLAN-0333-01, ANR-EMMA-002-01, ANR-14-CE26-0038, and ANR-14-CE32-0019, respectively), the iXCore Research Foundation, and the French government through its program ``Investments for the Future'' under the Universit\'e C\^ote d'Azur UCA-JEDI project (under the label Quantum$@$UCA) managed by the ANR (grant agreement ANR-15-IDEX-01).
P.V., C.B., E.G., H.H., C.S., O.A., S.T., and F.K. acknowledge the European Commission for the FP7-ITN PICQUE project (grant agreement No 608062). This work was also conducted within the framework of the project OPTIMAL granted by the European Union and the R\'egion PACA by means of the Fond Europ\'een de D\'eveloppement R\'egional, FEDER. R.N. and F.K. acknowledge support by the state of Baden-W\"urttemberg through the Center for Integrated Quantum Science and Technology (IQST).

We also thank A. Zavatta for help on the detection scheme using the femtowatt photoreceiver and P. Neumann for fruitful discussions.

\section{Additional information}
\textbf{Competing financial interests:} The authors declare no competing financial interests.

\textbf{Data availability:} Data is available from the corresponding author upon reasonable request.


\end{document}